\documentclass{spie}  
\usepackage[]{graphicx}

\def\arcsec{\hbox{$^{\prime\prime}$}}
\def\farcs{\hbox{$.\!\!^{\prime\prime}$}}
\def\degr{\hbox{$^\circ$}}

\title{AstraLux: the Calar Alto Lucky Imaging Camera} 

\author{Felix Hormuth, Stefan Hippler, Wolfgang Brandner, Karl Wagner, Thomas Henning
\skiplinehalf
	Max-Planck-Institut f\"ur Astronomie, K\"onigstuhl 17, 69117 Heidelberg, Germany
}

\authorinfo{Further author information: (Send correspondence to F. Hormuth)\\F. Hormuth: E-mail: hormuth@mpia.de, Telephone: +49 (0)6221 528 221}

\begin{document} 
  \maketitle 

\begin{abstract}
AstraLux is the Lucky Imaging camera for the Calar Alto 2.2-m telescope, based on an electron-multiplying high speed CCD. By selecting only the best 1-10\% of several thousand short exposure frames, AstraLux provides nearly diffraction limited imaging capabilities in the SDSS {\em i'} and {\em z'}\ filters over a field of view of 24$\times$24 arcseconds. \\
By choosing commercially available components wherever possible, the instrument could be built in short time and at comparably low cost. We present the instrument design, the data reduction pipeline, and summarise the performance 
and characteristics\footnote{Based on observations collected at the Centro Astron\'omico Hispano Alem\'an (CAHA) at Calar Alto, operated jointly by the Max-Planck Institut f\"ur Astronomie and the Instituto de Astrof\'isica de Andaluc\'ia (CSIC)}.
\end{abstract}



\section{INTRODUCTION}
The recovery of the full theoretical resolution of large ground-based optical telescopes
in the presence of atmospheric turbulence has been a major goal of technological
developments in the field of astronomical instrumentation in the past decades. 
The de-facto standard nowadays is adaptive optics, the active real-time correction of the incoming 
distorted wavefronts by means of deformable mirrors. While diffraction limited 
image quality can now be routinely achieved in the near infrared at large telescopes, this is 
only possible with considerable technical effort.

Even before the age of adaptive optics, speckle interferometric techniques  gave
access to the diffraction limit of large telescopes by applying auto- or cross-correlation 
methods in Fourier space to large sets of atmospherically degraded short-time 
exposures\cite{Labeyrie:1970,Labeyrie:1974,Knox:1973,Knox:1974,Lohmann:1983}.

An even more simple approach has been employed especially by amateur astronomers
in order to produce sharp images of the moon or bright planets: instead of using all images
of a large set of short-time exposures, only those showing little image distortion due to the
variable strength of atmospheric turbulence are
combined to a high-resolution result. Like in the speckle interferometric approach,
this requires only a detector with fast readout capability and moderate computational
effort. 
The application of this technique to fainter astronomical targets, e.g. low-mass double stars, was limited
by the readout noise of the available detectors at visible wavelengths until a few years ago. At typical exposure times of few ten
milliseconds -- necessary to ``freeze'' the effects of atmospheric turbulence -- photon noise
limited detectors are obligatory to use this so-called ``Lucky Imaging'' technique on targets fainter than $\approx$10\,mag. 
Since the image quality of each single frame has to be determined, e.g. by measuring the Strehl ratio of a 
reference star, the readout noise of the detector sets stringent limits on the minimum brightness of this reference.

The advent of electron multiplying CCDs (EMCCD)\cite{Jerram:2001,Hynecek:2002,Mackay:2001} led
to considerable interest in the Lucky Imaging technique on the part of professional astronomers. 
First experiments at the Nordic Optical Telescope (NOT) with LuckyCam\cite{Baldwin:2001,Tubbs:2002,Tubbs:2003,Law:2006} 
proved that Lucky Imaging is a very promising alternative to adaptive optics, allowing diffraction 
limited imaging at telescopes in the 2--3\,m class at wavelengths below 1$\mu$m.
These results triggered the development of a similar instrument for the 2.2-m telescope at
Calar Alto by our group. 
The Lucky Imaging camera AstraLux was built in less than 5 months thanks to the availability of most 
parts as off-the-shelf equipment. 

AstraLux routinely reaches Strehl ratios of
15--20\% in the SDSS {\em i'} and {\em z'} passbands and has become the standard
tool for double star observations at Calar Alto. Small technical overheads, an easy to use
software interface and the availability of pipeline results in near real-time allow the
observation of up to 20 targets per hour. 

In the following we describe the instrument design and data processing pipeline and
summarise the instrument's performance and key characteristics. Interested readers
looking for more comprehensive information are referred to the diploma thesis\cite{Hormuth:2007a}\footnote{Available online at \texttt{http://www.mpia.de/ASTRALUX/}}, which covers AstraLux and its performance in full detail.

\section{INSTRUMENT DESIGN}

\begin{figure}
	\begin{center}
   		\begin{tabular}{c}
   			\includegraphics[width=14cm]{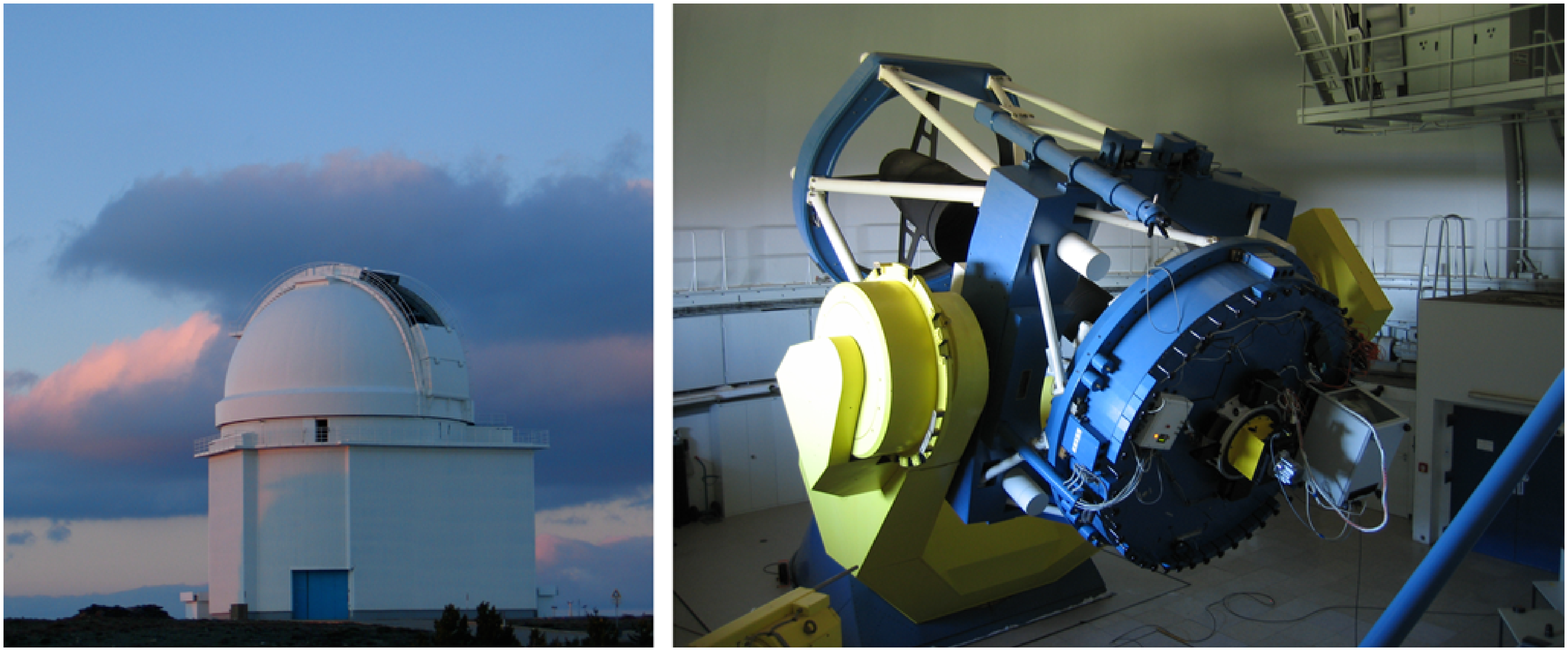}
			\\
			\includegraphics[width=14cm]{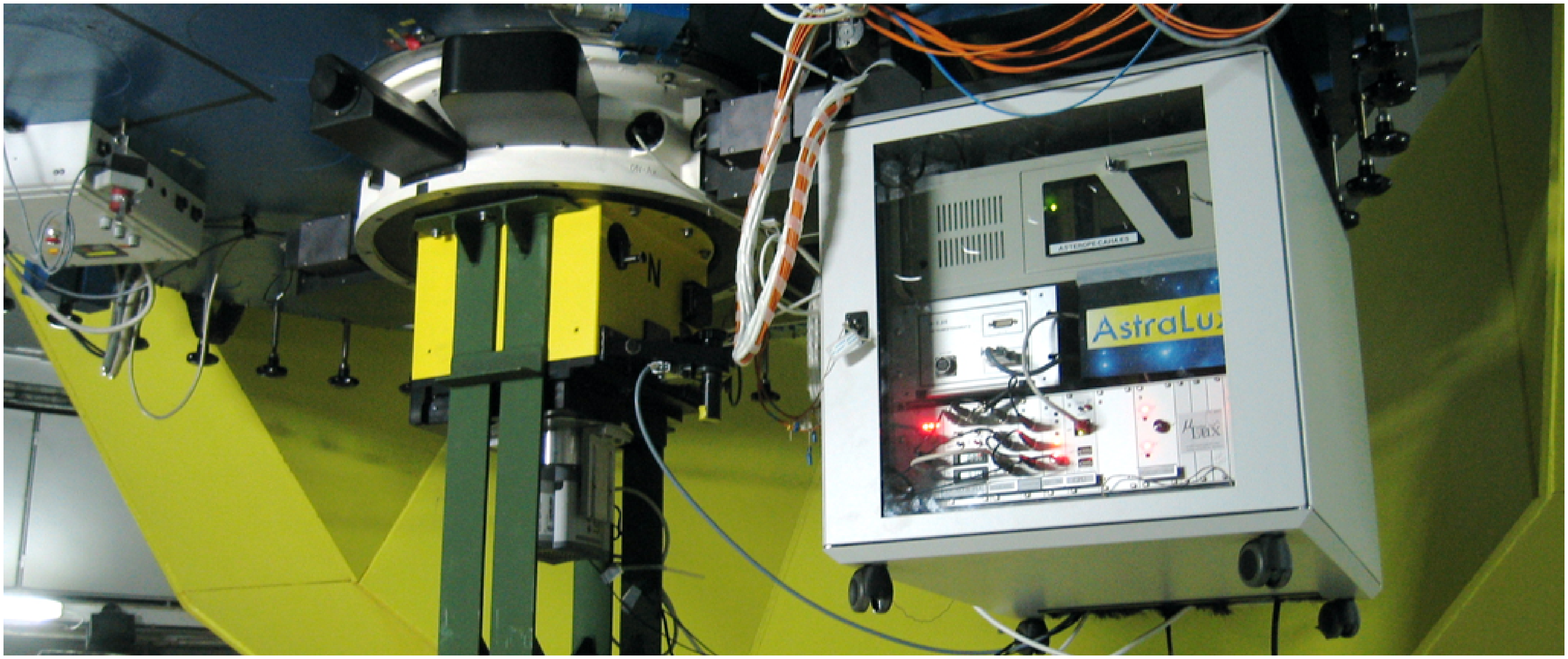}
 	 	 \end{tabular}
 	  \end{center}
  	\caption[example] 
		{ \label{fig:telescope} 
		AstraLux and the Calar Alto 2.2-m telescope. 
		
		{\em Top:} the 2.2-m telescope dome and the telescope
		with AstraLux attached at the Cassegrain focus. 
		
		{\em Bottom:} Detailed view of the instrument at the
		telescope. The yellow box contains the 8-position filterwheel, the camera is attached at the bottom.
		The grey rack to the right houses the camera control computer, a keyboard/monitor combination,
		the filterwheel electronics, and MicroLux, the GPS timing add-on.}
\end{figure} 
\subsection{Telescope and site characteristics}
The Calar Alto observatory is located at a height of 2150\,m in the Sierra de los Filabres, approximately 50\,km north
of Almeria in Andalusia, southern Spain. The median seeing, measured by the site's seeing monitor 
$\approx$300m away from the telescope is 0\farcs9, slightly better in summer than in winter\cite{Sanchez:2007}.

Together with the given telescope diameter of 2.2\,m, the seeing defines the optimal wavelength for
Lucky Imaging: the probability for the occurence of a good frame (i.e. with a Strehl ratio $>$37\%) in 
a series of short exposure times images decreases exponentially\cite{Fried:1965} with the ratio of telescope diameter
over Fried parameter $r_0$, which again depends inversely on the seeing and scales with $\lambda^{5/6}$. While observations at
short wavelengths result in too low probabilities for usable frames,  the gain in resolution by selecting only high-quality frames becomes small at
long wavelengths due to the
decreasing theoretical resolution of the telescope. For the Calar Alto 2.2-m telescope, the optimal wavelength
for efficient Lucky Imaging lies in the range of 800--1000\,nm, including the Johnson $I$ and the Sloan Digital
Sky Survey (SDSS) $i'$ and $z'$ passbands. At typical seeing conditions one can expect that $\approx$0.5--1\% of
all images will have an acceptable Strehl ratio of few 10\% while providing a 6 to 10 times better resolution.
A Lucky Imaging instrument for the 3.5-m telescope at the same site would have to operate optimally in the $H$-band -- a
wavelength range for which no photon noise limited detectors are available off-the-shelf so far.

\subsection{Camera \& Optics}
\subsubsection{Camera Head}
The AstraLux camera head is an Andor DV887-UVB model from Andor Technologies,
Belfast, Northern Ireland. It is an electron-multiplying, thinned, and back-illuminated 512$\times$512 
pixel CCD that comes as complete package with a multi-stage Peltier cooler, mechanical shutter, computer 
interface and software. 
It can be operated with readout clocks of up to 10\,MHz in frame transfer mode, allowing a 
full frame rate of 34\,Hz. 
Using subarrays, binning, and short vertical shift times allows 
frame rates of more than 1\,kHz. These short exposure times are realised by electronic shuttering,
whereas the mechanical shutter is only used for bias and dark frame acquisition and to
protect the CCD window against dust. 
Starting at room temperature, the typical operating temperature of $-$75\degr C is usually
reached within 10\,minutes. The camera requires neither refill of liquid coolants nor
any action to maintain the vacuum inside the CCD head.
The quantum efficiency (QE) of the E2V CCD97 detector peaks at $>$90\% in the $R$ band. At 912\,nm,
the central wavelength of the SDSS $z'$ filter, the QE is still $\approx$40\%.

The camera can be used either as conventional CCD without electron multiplication, or as EMCCD
with multiplication gains of up to 2500 at $-$75\degr C. The electron gain is set from within the control
software with a resolution of 255 steps, unfortunately not linearly related to the actual physical electron gain\footnote{
Newer models of Andor EMCCDs allow a direct, calibrated setting of the EM gain.}.
For Lucky Imaging we use a readout clock of 10\,MHz with 14\,Bit A/D resolution. The readout noise is $\approx$80e$^-$.
At an electron gain of 2500, this corresponds to a SNR of $>$30 for single photons. So called clock-induced charges (CIC),
spurious events caused during vertical clocking in the sensor or frame transfer area, occur with a probability of $\approx$5\%
at a vertical shift time of 1.8$\mu$s. The CIC probability strongly depends on the shift time and the applied shift
voltage. Both parameters can be varied via the control software, but it is not recommended to deviate from the
adopted standard settings. Not only the CIC probability, but also the charge transfer efficiency depends on these
parameters. At full electron gain, the camera is linear to up to 70 photons per pixel, or 170000 with disabled
electron gain.

\subsubsection{Barlow Lens}
At the Calar Alto 2.2\,m telescope, the camera's physical pixel size of 16\,$\mu$m roughly  corresponds
 to almost twice the size of the theoretical PSF at 912\,nm (SDSS $z'$ band). 
Diffraction limited imaging is thus not possible without some kind of magnification optics that 
increases the effective focal length. 
Proper sampling in the sense of the Nyquist criterion  can only be reached with a
magnification factor of 5 in SDSS $z'$, and even larger factors are necessary for observations 
at shorter wavelengths. 
However, for the final design, a value of 4 was adopted as a good compromise between spatial sampling 
and the size of the field of view which is 24$\times$24\arcsec\ at a pixel scale of 47\,mas/px.
The hardware realisation chosen for the AstraLux instrument is a single negative achromat in
Barlow configuration, i.e. placed in the optical path before the nominal focal plane of the telescope.

The AstraLux Barlow lens is a Thorlabs ACN127-030-B achromat with a diameter of $d$=12.7\,mm 
and a focal length of $f$=$-$30\,mm. 
It is optimised and coated for the wavelength range 650$-$1050\,nm. 
The lens provides diffraction limited image quality over the full field of view and the whole specified 
wavelength range. 
Field distortion of up to 0.5\% is the result of using a lens with such a short focal length in a 
non-telecentric configuration. 
Though this poses an instrumental limit to the astrometric accuracy if it is not properly calibrated, 
it allows a very compact and rigid construction of the fore-optics. \\
The lens is held in a standard C-mount tube --  bought ``off-the-shelf'' 
like the lens itself -- that is mounted directly on the camera. This guarantees a high stability of the
pixel scale, which strongly depends on the distance between lens and CCD. Tolerances for
tilt and de-centre of the lens were found to be 2\degr\ and 1\,mm, respectively, referring
to a geometric spot pattern size smaller than the diffraction limited PSF. 
The direct connection between camera and lens tube helps to achieve
a good alignment, and eases dis- and reassembly of the camera system.

\subsubsection{Filters}
For operations at the Calar Alto 2.2\,m telescope it was decided to refurbish the so-called
{\it Instrumenansatz\,1} (IA1), an adapter that was used for mounting conventional CCDs in the
past . 
This device can be mounted behind the video guider unit at the Cassegrain focus
of the telescope and includes a filter wheel with 8 positions. This wheel can hold
virtually any filter that is available at the observatory\footnote{See \texttt{http://www.caha.es//CAHA/Instruments/filterlist.html}
for a list of all available filters.}, allowing observations at a wide range of wavelengths. 
The filter wheel  control electronics is located in the electronics rack at the mirror cell, and
the wheel is controlled via a graphical user interface on the camera control computer. \\
The standard filters used for Lucky Imaging observations are SDSS $i'$ and $z'$ band interference
filters, manufactured by Asahi Spectra Co., Ltd., Tokyo, Japan. The filter transmission curves are very
close to the original SDSS filter system definition\cite{Fukugita:1996} and peak at more than 95\% transmission.

\subsection{Instrument Control}
AstraLux can be controlled from virtually any point of the observatory with a 100\,MBit Ethernet
network connection. The user connects via Remote Desktop Protocol (RDP) to a standard PC
workstation that serves as gateway to the camera control computer. This control computer, also
a standard PC, is located in the electronics rack at the mirror cell and exclusively used to run
the original camera control software supplied by the vendor and a simple user interface for
filter wheel control.\\
The camera control software offers a real time display and allows setting of all crucial parameters 
like exposure time, electron gain, or number of requested frames. The raw FITS data cubes (typically 18\,MB/s
at maximum frame rate) are not stored on the camera computer itself, but on a fast RAID-0 array
with 1\,TB capacity in the gateway machine.

\subsection{Data Processing \& Storage}
The AstraLux  data reduction software is running on a dedicated
pipeline computer, equipped with two dual-core Woodcrest processors and 8\,GB of memory.
Upon completion of an observation, the pipeline machine automatically fetches the raw
data from the primary storage RAID array and starts data processing. Telescope and environmental
parameters are retrieved from the Calar Alto database.
The pipeline then automatically produces quicklook results of the Lucky Imaging observations 
in approximately the same time that is needed for data acquisition. The basic pipeline algorithms are 
similar to that adopted by the LuckyCam team\cite{Tubbs:2003a,Law:2006a}.

At the end of the observing night, complete sets of raw data exist on the gateway machine as
well as on the pipeline computer. Backup is started in the morning, preferrably to an external USB
harddisk, to be supplied by the observer. Up to 400\,GB of raw data are produced in a typical
winter night, while the pipeline results need only $\approx$2\,GB of disk space.

\subsubsection{Calibration Data Processing}

Both bias and flatfield cubes are combined to master calibration images using
a kappa-sigma clipping algorithm. In the case of flatfield images acquired through the
conventional amplifier without electron multiplication, this will prevent cosmics from appearing in the final
product. Since bias frames are usually acquired with the same camera
parameters as the corresponding science observation, i.e. at high electron gains, 
they will most likely be contaminated by clock induced charges (CICs). 
With typically 50$-$100 single frames in a bias cube, these background events are 
 removed by kappa-sigma clipping as well.
All master calibration files are part of the final
set of pipeline products. 

\subsubsection{Lucky Imaging Data Processing}
First, the position of a suitable reference object for quality assessment has to be
determined. This is performed on a stacked image of the first 2 seconds of raw
data, and can be done either manually or automatically. While the manual
option allows to select any star just by clicking on the image, the automatic
reference finding algorithm will always choose the brightest object in the field of view.
The pipeline tracks the reference on all following raw data frames to account
for large atmospheric tip/tilt or telescope tracking errors.

\begin{figure}
	\begin{center}
   		\begin{tabular}{c}
		   	\includegraphics[width=12.5cm]{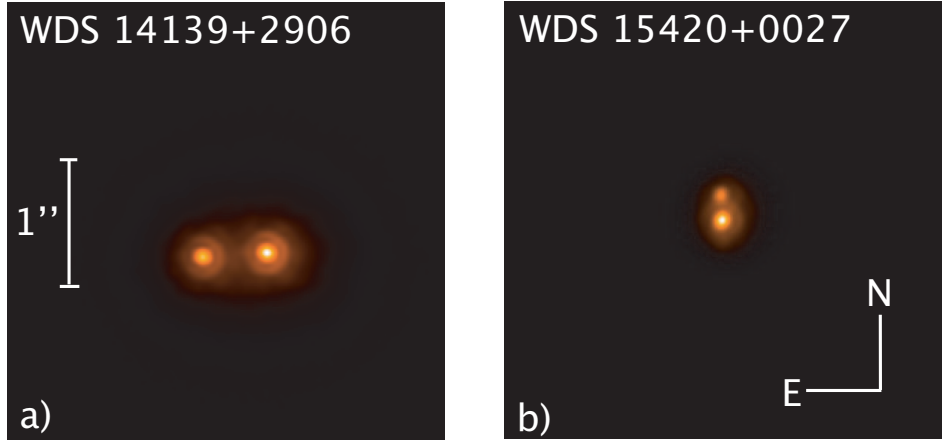}
 	 	 \end{tabular}
 	  \end{center}
  	\caption[example] 
		{ \label{fig:first22} 
AstraLux first light double stars: the binary WDS\,14139+2906 with an angular
		separation of 0\farcs52 and component magnitudes {\em V}=7.5 and {\em V}=7.6\,mag, and 
		the 0\farcs23 separated binary WDS\,15420+0027 with {\em V}=8.2 and {\em V}=8.8\.mag
		component brightnesses. Both images are based on a 2\%-selection from 10000 single frames
		in the SDSS {\em z'} filter with 30\,ms exposure time each. Image scaling is linear.
		\\
	}
\end{figure} 
\begin{figure}
	\begin{center}
   		\begin{tabular}{c}			
   			\includegraphics[width=15.2cm]{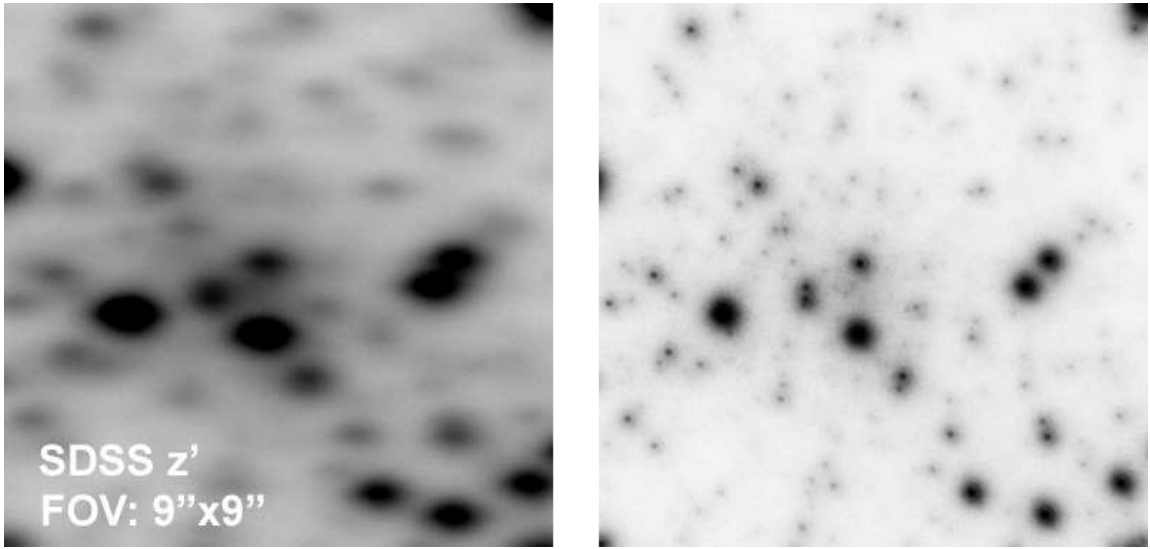}
 	 	 \end{tabular}
 	  \end{center}
  	\caption[example] 
		{ \label{fig:M15} 
	 Comparison between seeing limited imaging and the ``Lucky'' version: The combination of the best 5\% of 10000 single frames provided a Strehl ratio of 20\% in this observation of the core of the globular cluster M15. Though the conventional result contains 20 times more photons, it is clearly inferior in terms of point source detection limits. 
	}
\end{figure} 

Subsequently, the quality of each single frame is determined by measuring the Strehl ratio of
a reference object.
A small region with a size of typically 32$\times$32 pixel around the reference source
is extracted, resampled by a factor of four, and noise-filtered.
The resampling serves two purposes. 
Firstly, since the position of the brightest pixel of the reference object's PSF will be used as input to the
image reconstruction algorithm, sub-pixel shifting can only work if the reference position
is determined on resampled images. 
Secondly, the resampled images allow better estimates of the Strehl ratio. 
The Strehl value is derived from measurements of the ratio between peak
flux and total flux of the reference object -- a simple and fast method. 
Unfortunately, the peak flux in the slightly undersampled AstraLux images depends on the precise
position of the PSF peak within the brightest pixel, while the total flux is independent of the
reference object's position. 
Simulations with the AstraLux pixel scale of $\approx$47\,mas showed that Strehl measurements of a 
perfect PSF would suffer from a jitter of up to 20\%. 
This jitter can be reduced to 1\% and less by resampling the data before measuring the Strehl ratio. 
If the data is not only resampled, but also filtered with the telescope's modulation transfer function (MTF), 
the resampling will not introduce additional noise and single-pixel events like CICs and dark current electrons
can be suppressed. 

The pipeline's image reconstruction module performs data reduction in its literal sense. From
typically several GB of input data, just a few MB of pipeline results are produced. The most
interesting ones -- the Lucky Imaging results -- are currently generated with the Drizzle
algorithm\cite{Fruchter:2002}. This linear reconstruction method is flux preserving
and able to at least partially overcome the slight undersampling that is present in the raw data. 
It is capable of handling sub-pixel translations without the need to perform image shifting in the Fourier
domain. 
The current IDL implementation of the Drizzle algorithm is somewhat simplified and
does not consider image rotation or field distortions, but just shifts the selected images such
that the brightest pixel of the reference star is always positioned at the same pixel coordinates.
The drizzling process oversamples the input data twice, resulting in a pixel scale of $\approx$23.7\,mas/px in the final images.
Currently the pipeline produces drizzled results of the best 1, 2.5, 5, and 10\% of the input frames.
Bias and flatfield calibration frames are applied to the input images prior to drizzling.
A seeing (and tracking) limited image  is generated as well using all frames to allow quick measurements of the seeing conditions, useful at times when the observatory's seeing monitor is switched off, or for later
assessment of the data quality.

\begin{figure}[tb]
	\begin{center}
   		\begin{tabular}{c}
   			\includegraphics[height=9cm,angle=270]{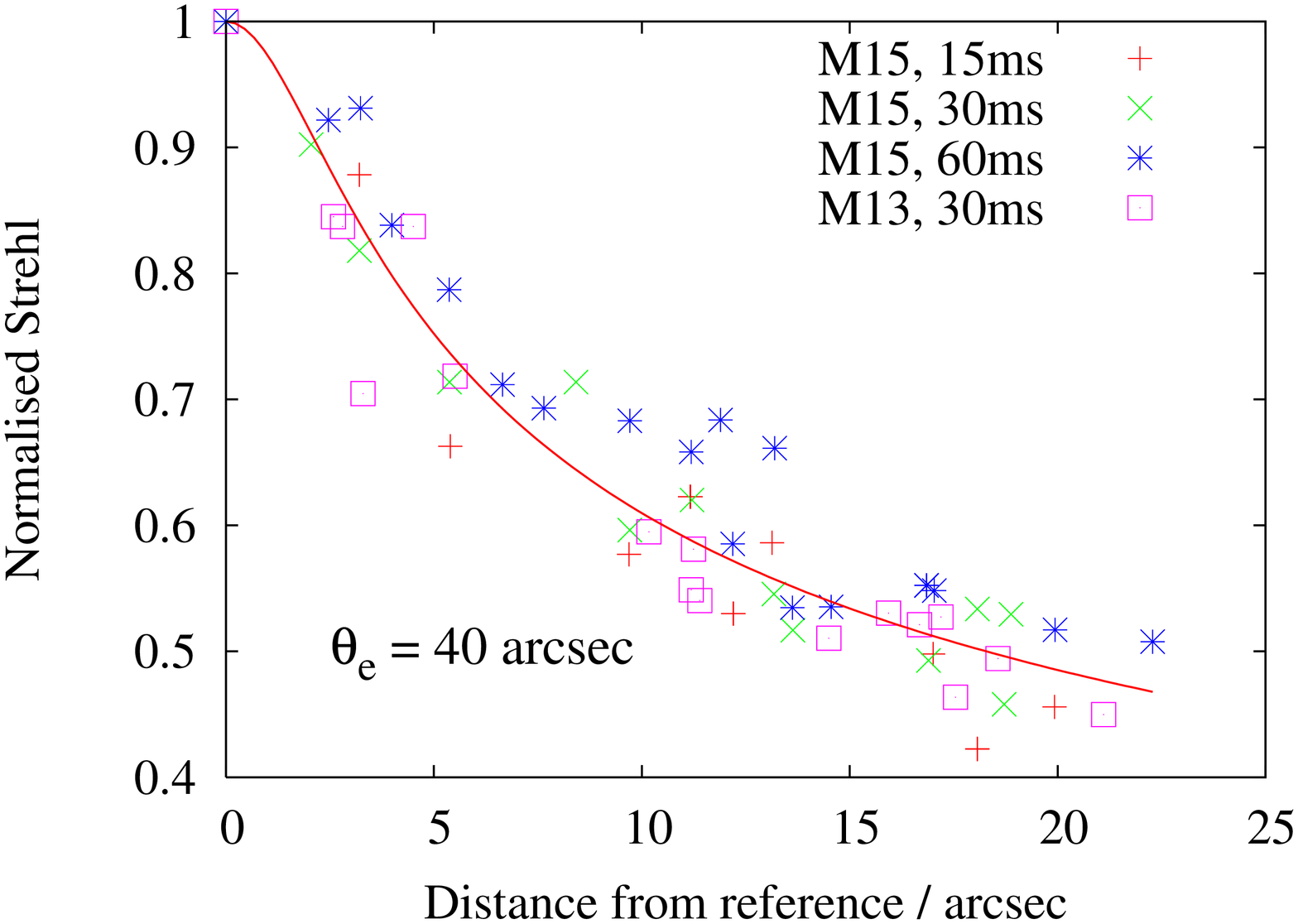}
   			\includegraphics[height=8.7cm,angle=270,trim=0cm 0cm 0cm 1cm]{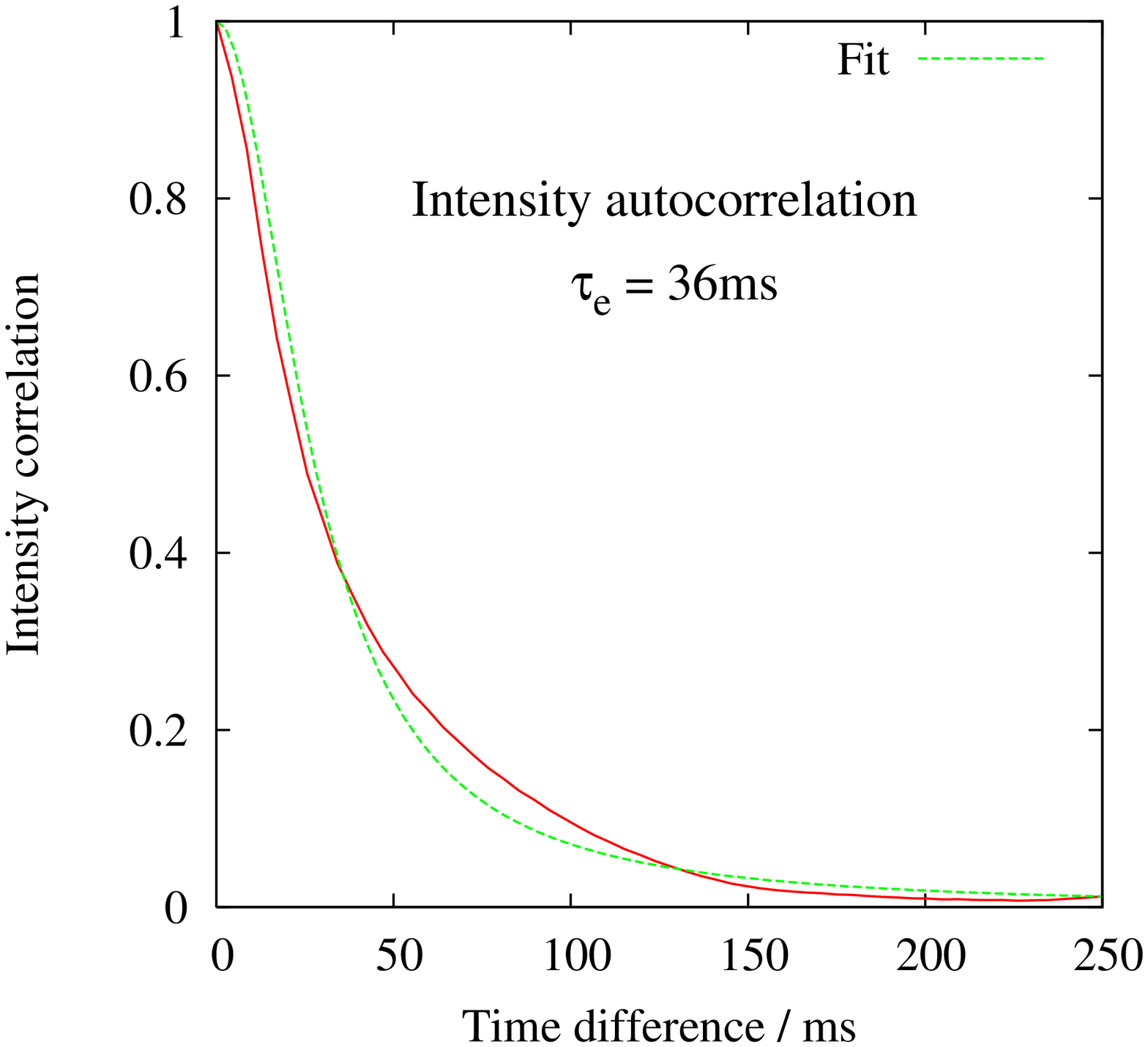}	
 	 	 \end{tabular}
 	  \end{center}
  	\caption[]{
		\label{fig:isoplan} 
		{\em Left:} Normalised Strehl ratio versus angular separation from the reference star for
		a range of single frame exposure times. All values are derived from globular cluster
		centre observations in the SDSS {\em z'} filter with 1\% image selection rate. The given
		isoplanatic angle of $\theta_e=$40\arcsec\ was derived by fitting a Moffat profile (solid line).
		
		{\em Right:} Normalised temporal autocorrelation function of the focal plane intensity. The 
			data was fitted with a Lorentzian profile, resulting in an estimated 
			speckle coherence time of $\tau_e$=36\,ms.
		}
\end{figure} 

\begin{figure}[tb]
	\begin{center}
   		\begin{tabular}{c}
			\includegraphics[height=8cm,angle=270]{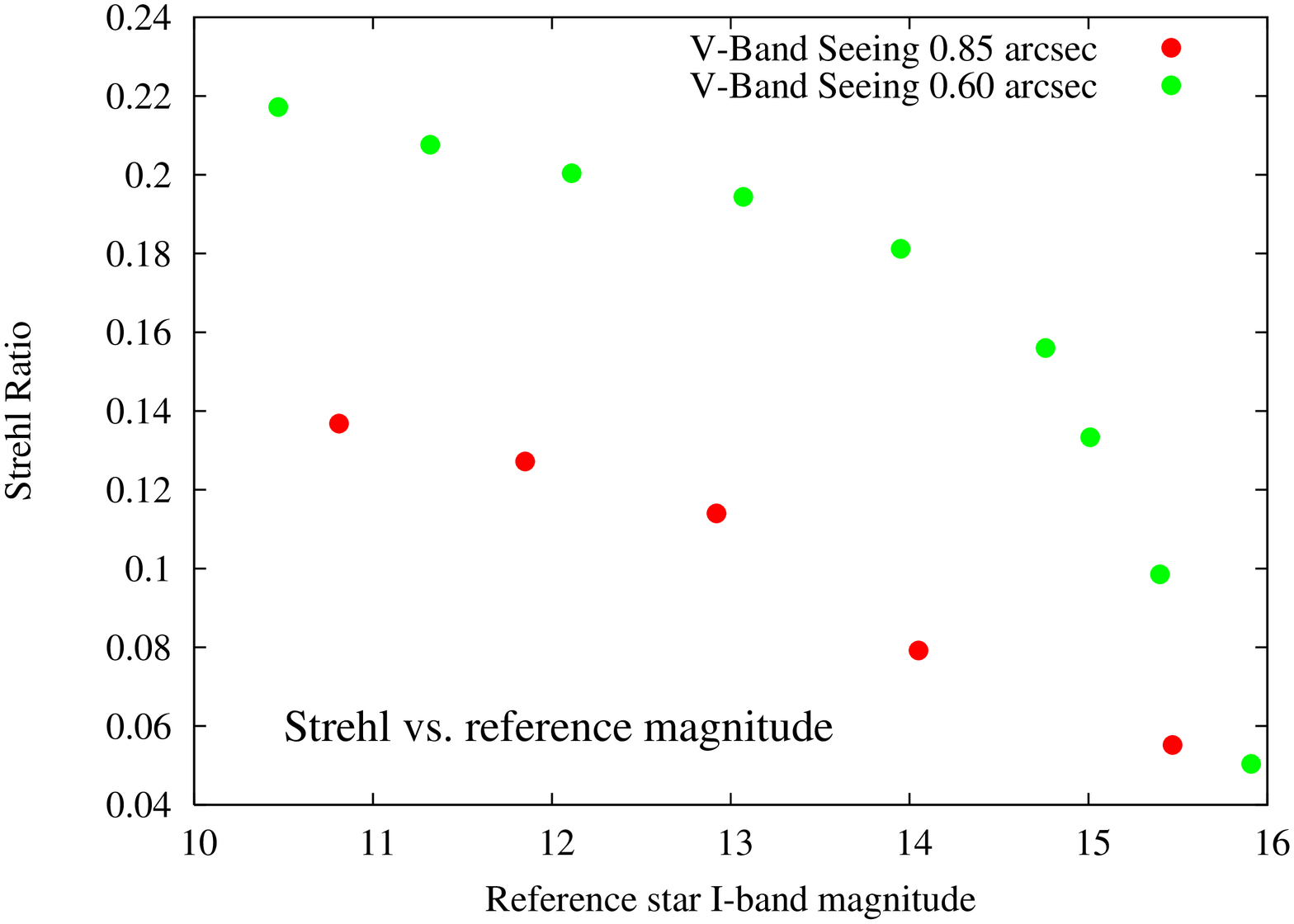}
			\hspace{1cm}
   			\includegraphics[height=8cm,angle=270]{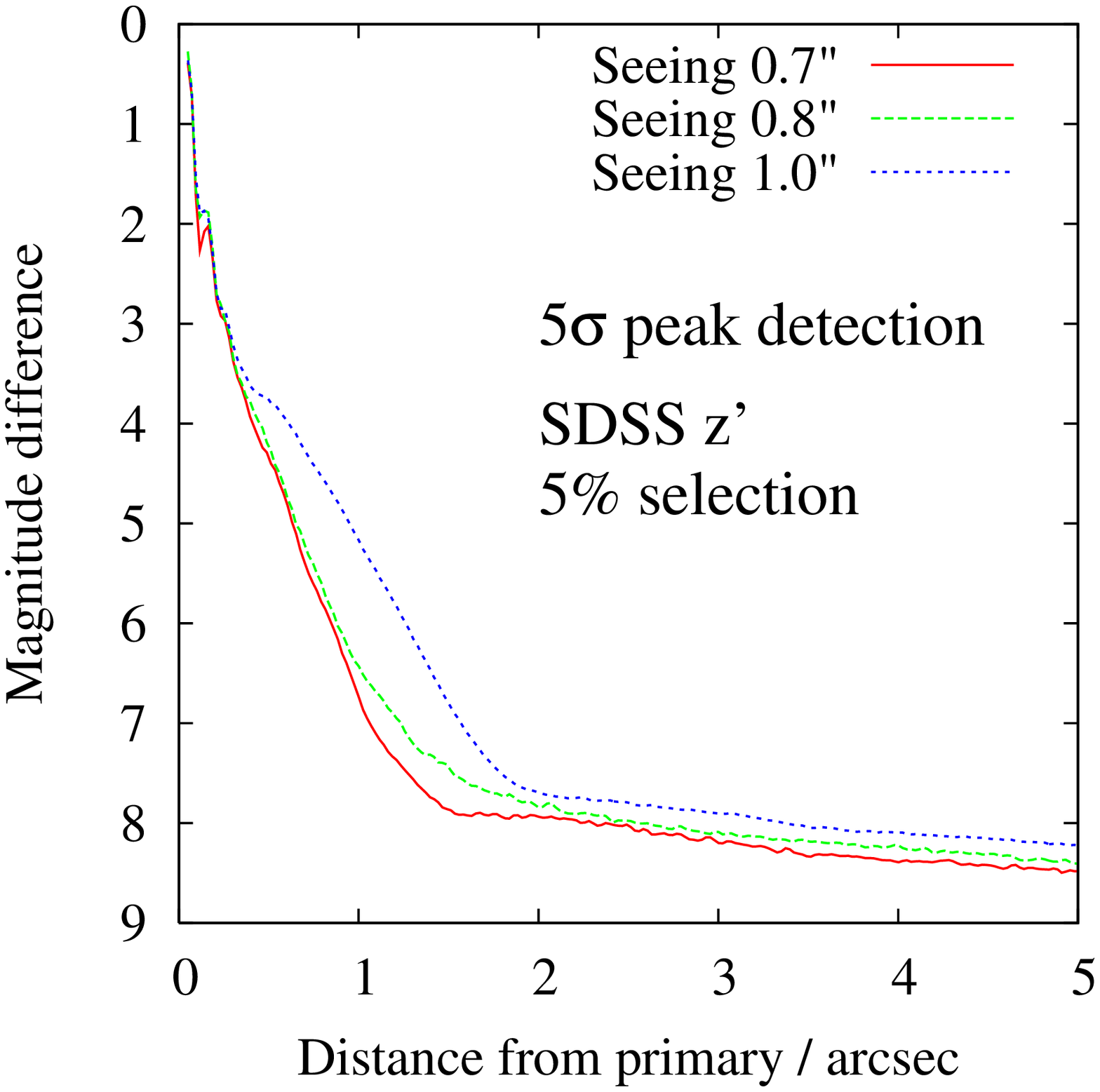}
 	 	 \end{tabular}
 	  \end{center}
  	\caption[example] 
		{ \label{fig:magdiff} 
		{\em Left: } Dependency of the final Strehl ratio on natural seeing and reference star magnitude.
		The plot is based on observations with 30\,ms single frame exposure time in the SDSS {\em z'}		
		filter and 1\% image selection rate. All Strehl ratios were measured on stars with less than 2\arcsec\
		separation from the reference star.	
		
		{\em Right:} Achievable magnitude differences for a 5$\sigma$ peak detection of a fainter companion
		to the reference star. All curves refer to a 5\% selection rate from 10000 images with 30\,ms single
		frame exposure time, equivalent to an effective integration time of 15\,s. The {\em I}-band magnitude
		of all three reference stars was $\approx$10\,mag.}
\end{figure} 

\section{SYSTEM PERFORMANCE}
First light at the Calar Alto 2.2-m telescope was obtained on July 6, 2006. From a total
of 5 nights during the first observing run, only half a night was lost due to bad weather, and
the remaining time provided photometric sky conditions with {\em V}-band seeing values 
as low as 0\farcs6.

Known bright double stars were chosen as first light targets and observed in the SDSS {\em z'}\ 
band with resulting Strehl ratios of $\approx$20\% (see Figure\,\ref{fig:first22}).
Operating the instrument, and especially acquiring targets, proved to be much easier than anticipated.
Though the pointing accuracy of the telescope is in general not better than 10\arcsec, the availability of
the camera's real time display allows short acquisition times of typically 1--2\,min per target.

AstraLux observations of globular cluster centres enabled the characterisation of the image quality
over the full field of view. Choosing different stars with a wide
range of magnitudes as the Lucky Imaging reference allowed to estimate brightness limits
for the reference selection and to measure the dependency of the Strehl ratio on the reference magnitude.
Among the globular clusters M3, M13, and M15, the latter  has been observed most extensively
with AstraLux. Figure~\ref{fig:M15} shows a comparison of a Lucky Imaging result to the corresponding
seeing limited image.

Astronomical observations with AstraLux started right during the very first night at the telescope, leading
to two refereed publications so far\cite{Hormuth:2007,Sicilia:2008}.
Completed and ongoing observing programs cover a wide range of target types, e.g. young stars
in nearby moving groups, low-mass binaries, T\,Tauri binaries, transition disk objects, minor planets, and 
microquasars. 


\subsection{Isoplanatic Angle and Coherence Time}
In the following, the isoplanatic angle $\theta_{\rm e}$ is defined as the angular separation from the
reference star, where the observed Strehl ratio drops to the fraction 1/$e$ of the reference star's Strehl ratio.
The observations of centres of the globular clusters M13 and M15 allowed measurements of the Strehl ratio for
a large number of stars, well distributed over the field of view. Figure~\ref{fig:isoplan} shows
the normalised Strehl ratios, i.e. the measured Strehl ratios divided by the reference Strehl, 
for three different M15 and one M13 observation. The images used for this plot are based
on a 1\% selection from 10000 frames in the SDSS\,$z'$ band. M15 was observed with 
three different single frame exposure times, without a significant impact on the {\em normalised}
Strehl ratios. The {\em absolute} Strehl ratio slightly decreases with increasing exposure time, though,
dropping from $\approx$22\% at 15\,ms to 18\% at 60\,ms.

A Moffat profile was fitted to the data, allowing extrapolation to the 1/$e$ point, and hence an estimate of the isoplanatic angle 
$\theta_{\rm e}$. 
For the SDSS $z'$ band observations, $\theta_{\rm e}$ has been found to be $\approx$40\arcsec, 
larger than the diagonal field of view size of 34\arcsec.

This measurement has been repeated for an observation of M15 in the Johnson $I$ band.
No significant difference to the isoplanatic angle extrapolated from the $z'$  band 
data was found. Theory predicts a decrease of $\theta_{\rm e}$ by a factor 1.05 when switching from 
$z'$ to $I$-band, too small to be reliably detected in the available data.
The expected isoplanatic angle in $V$-band is $\approx$2-5\arcsec\,
equivalent to $\approx$4-10\arcsec\ in the $z'$-band, but the isoplanatic angle in speckle observations
is typically about 1.7 times larger\cite{Vernin:1994}.
The AstraLux data and comparable measurements at the NOT\cite{Tubbs:2003a} suggest that the
selection process of the Lucky Imaging technique can further increase $\theta_{\rm e}$ to values as large
as 40$-$50\arcsec.

To determine the speckle pattern coherence time, a series of 10000 images of the bright star $\beta$\,{\em And} 
was recorded with a time resolution of 4.6\,ms. The normalised  temporal autocorrelation function is shown in 
Figure\,\ref{fig:isoplan} (right side). The measured coherence time $\tau_e$=36\,ms corresponds to the
1/$e$ point of a Lorentzian fit.


\subsection{Limiting Magnitudes}
The M13 and M15 observations were re-analysed with different choices of the Lucky Imaging
reference star to assess the impact of the reference magnitude on the final Strehl ratio.
Figure~\ref{fig:magdiff} shows the results for measurements under two different seeing conditions.
While reference stars as faint as $I$=15.5\,mag still allow a substantial improvement of image
quality under a 0\farcs65 seeing, the same performance cannot be reached with stars fainter
than 13.5\,mag in 0\farcs85 seeing. 

Another important number, especially in the context of binarity surveys, is the achievable contrast
ratio or magnitude difference for close companions to bright host stars. 
This parameter was determined on final pipeline results of SDSS\,$z'$ band observations
under different seeing conditions. All observed stars had an $I$-band magnitude of $\approx$10\,mag.
The achievable magnitude differences for a 5$\sigma$ peak detection are based on measurements
of the noise in concentric rings around these stars. This method suffers from the low number
of available pixels in the innermost 100\,mas, but is quite robust at larger angular separations.
Simulations with observed PSFs were carried out to check the reliability of the numerical results. \\
Figure~\ref{fig:magdiff} shows typical detection limit plots for three different $V$-band seeing
values. 
At angular separations larger than 2\arcsec, the
detection limit is determined by readout noise and the Poisson noise of the sky background.
Using more input images, i.e. increasing the effective exposure time of the Lucky Imaging result,
will increase the maximum achievable magnitude difference at large separations.
  \begin{figure}[b]
	\begin{center}
   		\begin{tabular}{c}
			\includegraphics[width=4.6cm,angle=270]{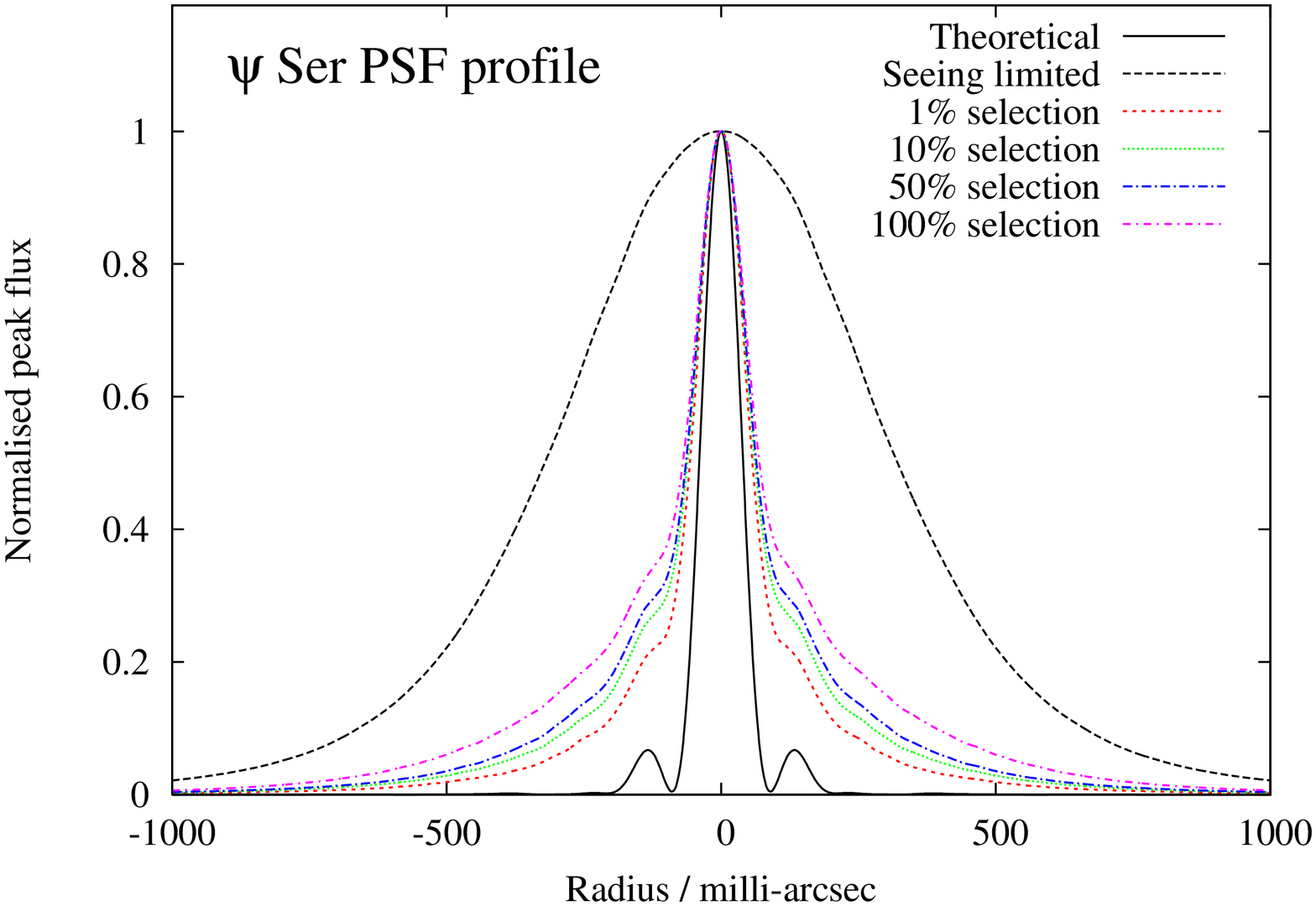}
			\hspace{1cm}
   			\includegraphics[width=4.6cm,angle=270]{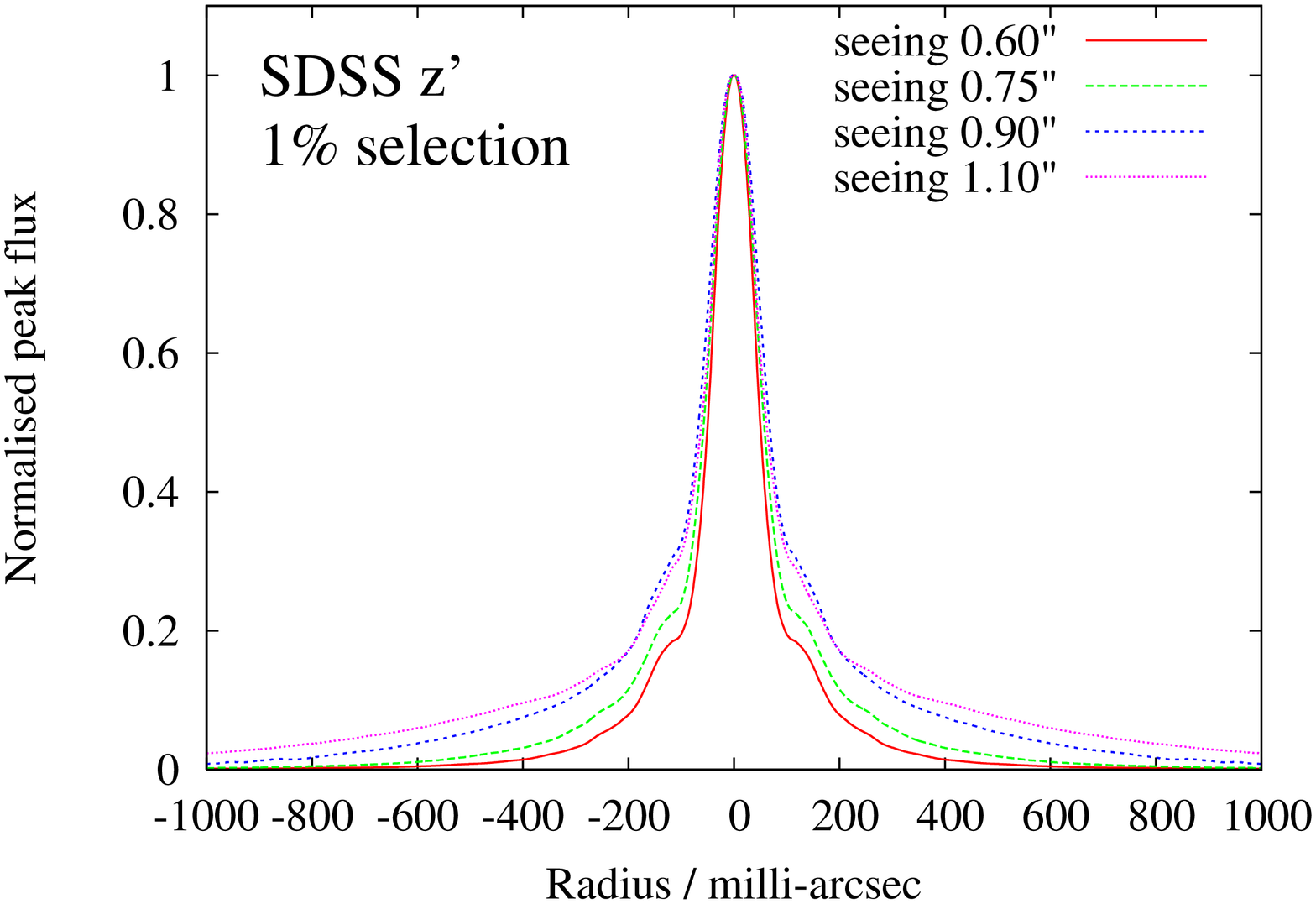}
 	 	 \end{tabular}
 	  \end{center}
  	\caption[example] 
		{ \label{fig:profiles} 
		{\em Left: } Radial profiles of the stellar PSF for different image selection rates under 0\farcs75
		V-band seeing. The 
		seeing limited and diffraction limited PSF are plotted for comparison. All 
		profiles have been normalised to a peak flux of 1. The resulting Strehl ratios for 1, 10, 50, and 100\%
		selection are 14.2, 10.4, 7.7, and 6.2\%. The seeing limited image has a Strehl ratio of 1.9\%.
		The FWHM of the Lucky Imaging PSF core ranges from 114 to 138\,mas.
		
		{\em Right:} Radial PSF profiles for different seeing conditions. All profiles refer to
			1\% selection from $\ge$10000 images with 15\,ms single frame exposure time.
			Strehl ratios for V-band seeing values of 0\farcs60, 0\farcs75, 0\farcs90, and 1\farcs10
			are 22.4, 14.2, 7.1, and 5.2\%. The FWHM of the PSF core ranges from 98 to 122\,mas.
		}
\end{figure} 

\begin{figure}
\begin{centering}
	\hspace{5mm}
	\includegraphics[width=4.6cm,angle=270,trim=0cm 3cm 0cm 3cm]{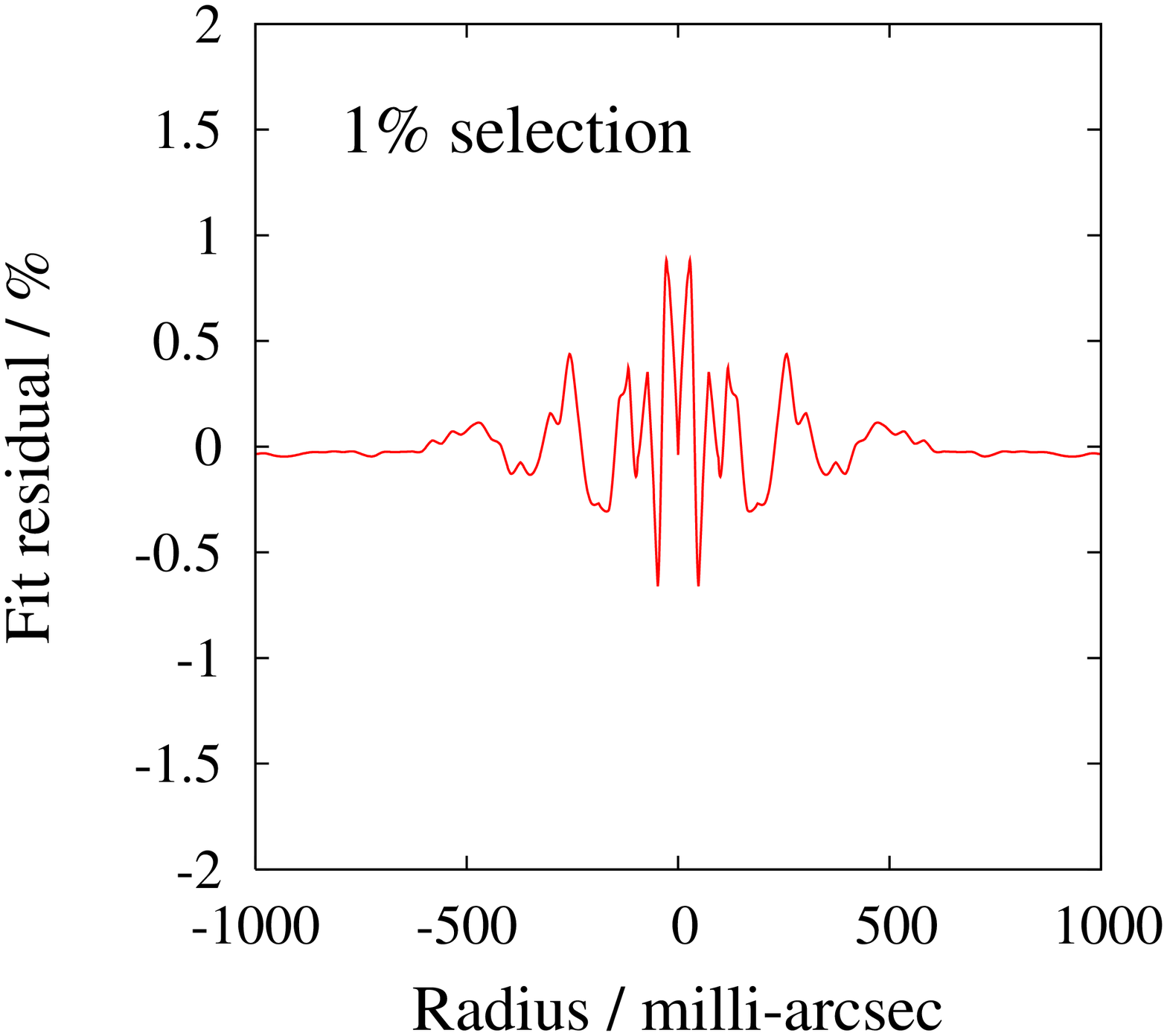}
	\includegraphics[width=4.6cm,angle=270,trim=0cm 3cm 0cm 3cm]{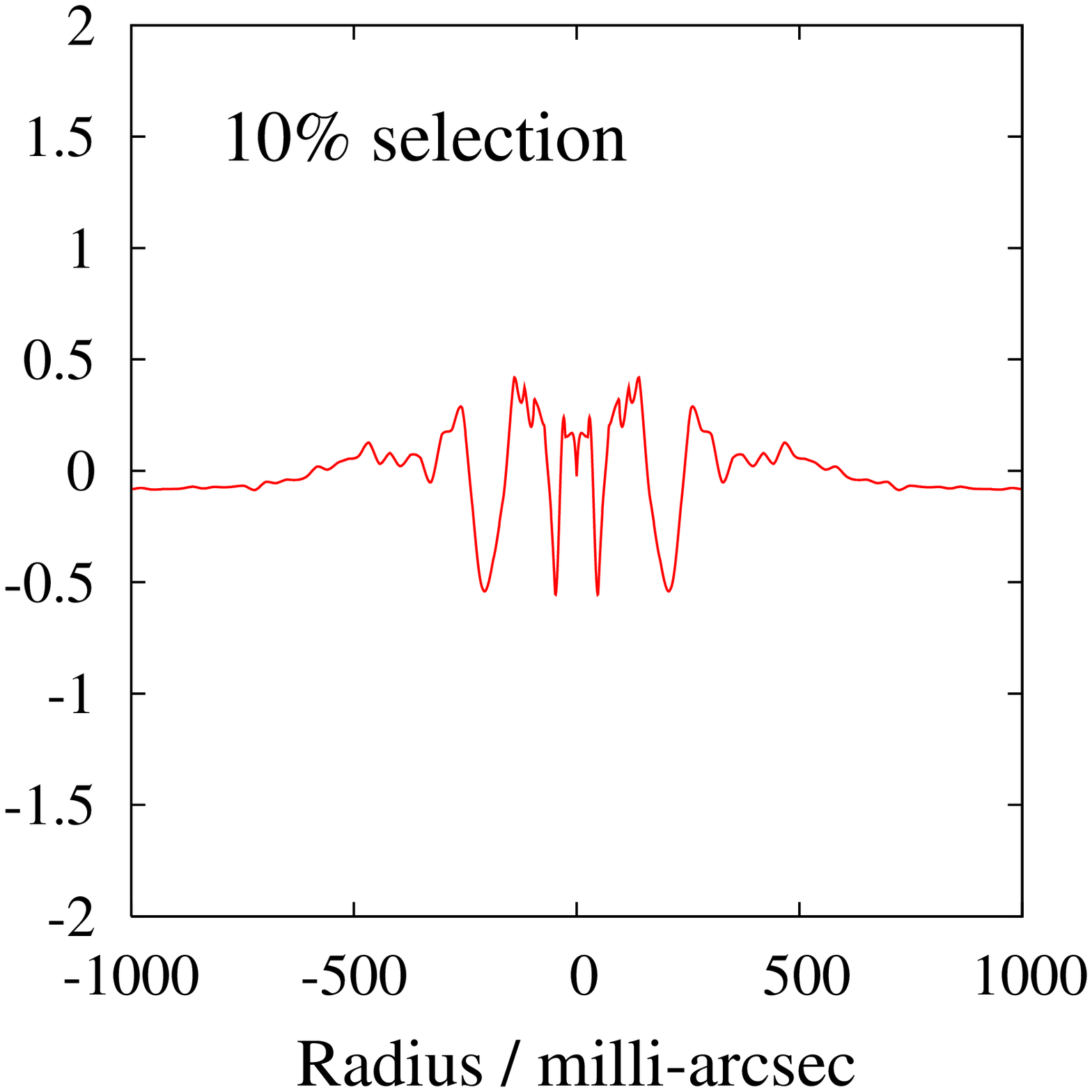}
	\includegraphics[width=4.6cm,angle=270,trim=0cm 3cm 0cm 5cm]{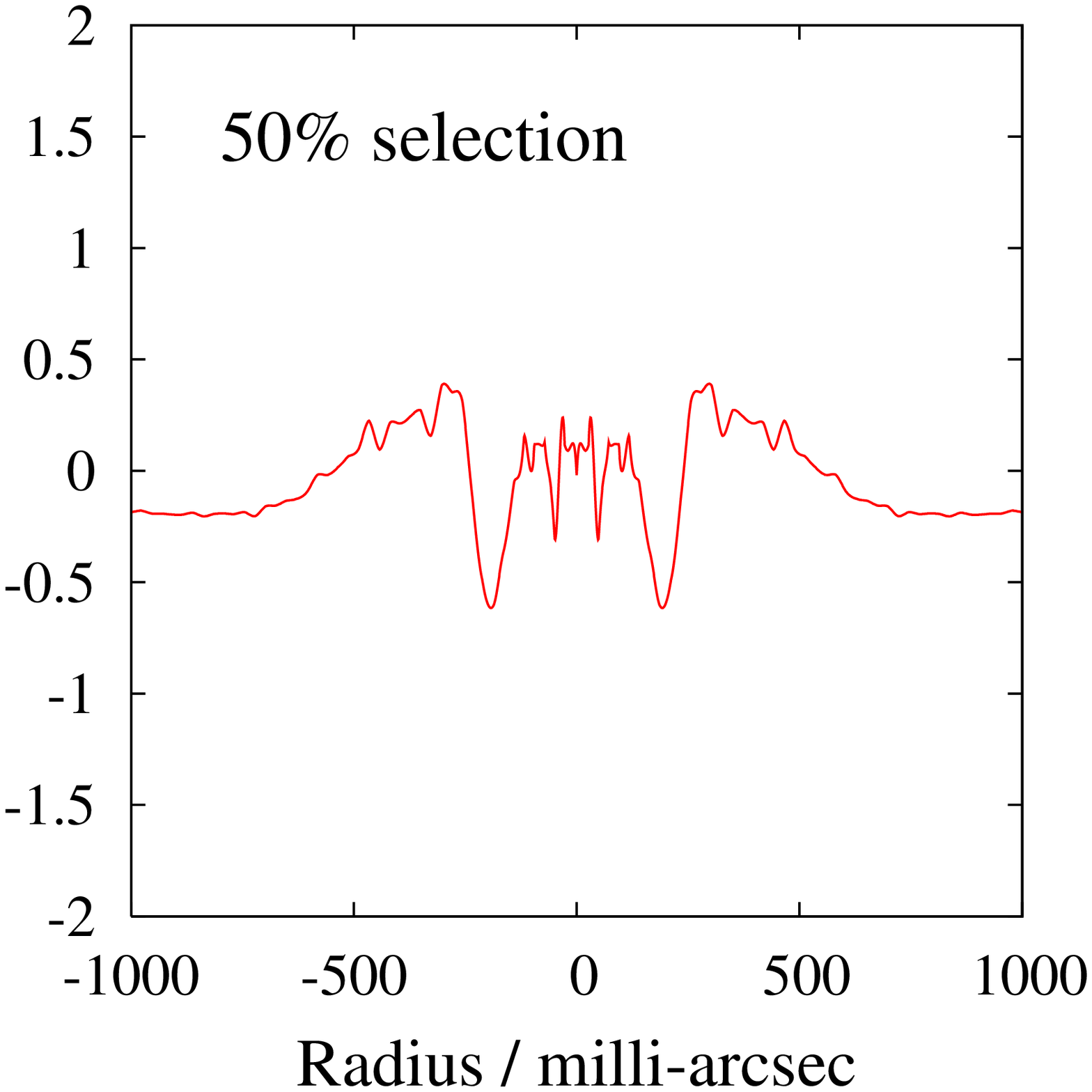}
	\vspace{2mm}
	\caption[]{Fit residuals for the application of the model described by Equation~\ref{eq:fitprof} to
		the radial profiles of the $\psi$\,{\em Ser} observations. The
		residuals are given in percent of the peak flux value.
		\label{fig:fitresid}}	
\end{centering}
\end{figure}

\subsection{Point Spread Function Characteristics}
In the absence of atmospheric dispersion effects, the Lucky Imaging point spread function
(PSF) of the AstraLux instrument shows a remarkable radial symmetry\footnote{Only in one observing
run in 2007 we observed static coma-like aberrations, probably related to a position-dependent tilt
of the primary mirror.}. There are no long-lived speckles 
like they are frequently observed in adaptive optics images. Using a sufficiently large number of input images -- typically
several thousand -- efficiently averages out asymmetries. For the quantitative assessment of the PSF
shape, it is therefore sufficient to consider the radially averaged profile only.

The left side of Figure~\ref{fig:profiles} shows the radial PSF profile of $\psi$\,{\em Ser}, observed
under a $V$-band seeing of 0\farcs75 with 15\,ms single frame exposure time through
the SDSS $z'$ filter. 
The profile is plotted for a range of selection rates from the 10000 input images. The theoretical and seeing limited PSF profiles are overlayed for comparison.
The profile of the theoretical PSF was derived from a simulated diffraction
limited image
with 2.237\,mas/px pixel scale, whereas the raw observational data was sampled with $\approx$46.6\,mas/px. Though the Drizzle process of the pipeline resulted in a final pixel scale of 23.3\,mas/px, principally
providing proper sampling in the sense of Nyquist, this is not sufficient to reconstruct a perfect PSF
even in the absence of any aberrations due to atmospheric turbulence or telescope imperfections.
However, for the judgement of image quality and the measurement of Strehl ratios, only the 
perfectly sampled PSF will be considered throughout this chapter. 

To calculate the Strehl ratios, all radial profiles were interpolated on a common two-dimensional
radius grid and numerically integrated to derive the total flux. 
The Pseudo-Strehl, i.e. the ratio of peak flux over
total flux, was calculated for each profile, and divided by the Pseudo-Strehl of the theoretical
PSF to obtain the real Strehl ratio. Since the radial profile of a source can be reliably 
reconstructed in the presence of a nearby companion or in a crowded field, this
method results in more robust estimates of the total source flux -- and hence Strehl ratio -- 
than standard aperture photometry. 
	
An increase of the selection rate causes only 
moderate broadening of the PSF core's FWHM, but a stronger decrease of the
resulting Strehl ratio due to more pronounced PSF wings. 
Even at only 1\% selection rate, the FWHM is considerably larger than theoretically expected. 
This is probably a result of the slight undersampling of the
raw data. Simulations have shown that with the current
setup and pipeline algorithms, the best FWHM to be expected  is $\approx$95\,mas. 

In a next step, the radial profiles of single stars were extracted from observational data over a 
range of seeing conditions. While a Strehl ratio of more than 15\% can be reached under a $V$-band
seeing of 0\farcs75 and better, it rapidly drops to a few percent if the seeing
gets worse than 1\arcsec. The radial profiles for a selection rate of 1\% in the 
SDSS $z'$ band and a single frame integration time of 15\,ms are plotted in
Figure~\ref{fig:profiles} for four different seeing values. As in the case of larger selection rates, an increase of the natural seeing leads in the
first instance to more pronounced PSF wings and has only moderate effects on the FWHM of the 
PSF core.

\subsection{PSF Modelling} 
The additional broadening of the Lucky Imaging PSF core and the wings of the radial profiles can
be modelled quite accurately as the weighted sum of a broad Moffat profile and a theoretical PSF that was convolved with a Gaussian: 

\begin{equation}
	\label{eq:fitprof}
	{\rm PSF}_{\rm obs}(r) = W \left(\frac{1}{r^2/\sigma_{\rm m}^2 +1}\right)^\beta + (1-W) \left({\rm PSF}_{\rm th}(r) \ast \exp\left( -\frac{r^2}{2\sigma_{\rm g}^2}\right)\right) 	
\end{equation}
 
Here, W weights the two PSF components, $\beta$ is the Moffat power law index, and $\sigma_g$
and $\sigma_m$ define the widths of the Gaussian and Moffat profile, respectively. PSF$_{\rm obs}$ and
PSF$_{\rm th}$ refer to the observed and theoretical radial PSF profiles.
This semi-analytical model has been applied to the $\psi$\,{\em Ser} data presented above. 
The resulting  fit parameters are given in Table~\ref{tbl:fitser}, whereas Figure~\ref{fig:fitresid} shows the  
residuals.
\begin{table}[htb!]
	\center
	\caption{ Semi-analytical PSF fit parameters for the $\psi$\,{\em Ser} observations.
		\label{tbl:fitser}}
	\begin{tabular}{c c c c c}
	\noalign{\smallskip}
	\hline\hline
	\noalign{\smallskip}
	Selection rate	& W & $\sigma_{\rm g}$ [mas] & $\sigma_{\rm m}$ [mas] &$\beta$\\
	\noalign{\smallskip}
	\hline
	\noalign{\smallskip}
	
	1\%	&		0.25	&	23.8	& 	250	&	1.62 \\
	10\%	&		0.31	&	24.4 & 	270	&	1.61 \\
	50\%	&		0.36	&	24.7	&	300	&	1.63 \\	
	
	\end{tabular}
\end{table}	

While the weighting factor and the width of the Moffat profile vary considerably under
changing selection percentages, the Moffat power law index and the width of the
Gaussian convolution kernel keep nearly constant. The possibility to reconstruct the
observed PSF profile from the known theoretical PSF and only two model parameters
is particularly interesting for binary fitting and PSF subtraction applications. In larger fields, the
dependency of the model parameters on the source position could be determined,
allowing PSF based photometry in crowded fields like globular cluster centres. \\
Figure~\ref{fig:fitresid} shows that the available theoretical PSF does not reproduce the
second diffraction ring correctly, leaving residuals of up to 0.5\% at $\approx$230\,mas distance from
the PSF centre. This is possibly related to the method used for PSF simulation. Currently, the theoretical
PSFs are computed for a circular aperture with central obstruction by the
secondary mirror, but do not include any contributions from the secondary spider
or any optics behind the primary mirror. 

The described model has been applied only in its radial symmetric form so far.
Further developments could include asymmetries due to atmospheric dispersion,
and it might be investigated if and how accurately the model parameters can be 
predicted based on the seeing-limited PSF only.

\subsection{Performance summary}
AstraLux is able to reach Strehl ratios as high as 25\% at the Calar Alto 2.2-m telescope
in the {\em z'} band. While this is only possible under a superior seeing of 0\farcs6 or better, the
median Calar Alto seeing of 0\farcs9 still allows to achieve Strehl ratios of more than 10\%.
The typical seeing limited Strehl ratio in SDSS {\em z'} under such conditions is $\approx$1.1\%.
In general, Lucky Imaging provides an improvement of the Strehl ratio by a 
factor of 10, corresponding to an increase of the signal-to-noise ratio for point sources by
a factor of 10$-$20, depending on atmospheric conditions.
Thus a selection of only the best 5$-$10\% of all images does definitely not have a negative effect on
the detection limit for point sources.

The requirements for the reference star magnitude are similar as for observations with 
adaptive optics.
The performance starts to significantly decrease at $I$=14\,mag, but 
image quality improvements are still possible with stars as faint as 15$-$16\,mag. 
The measured isoplanatic angle in $I$-band is with $\approx$40\arcsec\ as large as 
typical values in $K$-band for adaptive optics observations. 
Lucky Imaging performs considerably better than speckle imaging techniques. 
The typical magnitude limit for these methods is $V$$\approx$12\,mag, and the isoplanatic 
angle in $I$-band is only half as large.

The measured close companion detection limit at an angular separation of 1\arcsec\ is
on average 6\,mag. 
This is worse compared to the performance of AO systems --  here 
usually 8$-$10\,mag are reached and even more can be expected from future
AO systems with larger actuator numbers.
But: adaptive optics at 8-m class telescopes currently provides this capability only in the $H$ and $K$-band at 
wavelengths $>$1.5\,$\mu$m.
The achievable contrast ratio in speckle imaging observations is typically two 
magnitudes less than for Lucky Imaging.

Single frame exposures are limited to integration times below the speckle coherence time
of typically 30--40\,ms in SDSS $z'$. Like the size of the isoplanatic angle, this number shows seasonal
variations and depends on the natural seeing conditions.

Our measurements indicate that AstraLux is comparable to LuckyCam
at the Nordic Optical Telescope in terms of limiting magnitude and isoplanatic angle. 
Detector, optics, electronics, and software worked as
expected. The simple design of the instrument certainly contributed to a
smooth and satisfactory start of observations at Calar Alto.

\section{CONCLUSIONS}
Within less than one year it was possible to design, build, and characterise a Lucky Imaging
instrument for the Calar Alto 2.2-m telescope. Beyond evaluation of the observing technique
and data reduction strategies, scientific observations commenced right in the first night
at the telescope. As a common user instrument from 2007 on, AstraLux has become the standard 
tool for diffraction limited imaging at the Calar Alto observatory. It is currently mostly used for
binarity surveys among stars and minor planets. 

The encouraging results regarding both development speed as well as scientific output
triggered the start of the project ``AstraLux Sur'', an almost identical copy of the Calar Alto
version to be used as visitor instrument at ESO's New Technology Telescope (NTT) at La Silla, Chile.
AstraLux Sur is currently on its way to Chile after a development time of less than 2 months and 
will have first light in mid-July 2008.

Apart from Lucky Imaging, AstraLux can be used as a high speed photometer with single
photon detection capability. In combination with the custom GPS timing hardware MicroLux\cite{Hormuth:2008}
it has been successfully used for observations of the optical pulse profile of the Crab pulsar.

We are currently investigating possible improvements of the optical design, namely a telecentric
magnification optics and a tunable atmospheric dispersion corrector. However, the strength of
the current system is its simple design, leading to short development times, low hardware costs,
and high stability. \\
An ambitious project for the future would be the combination of AstraLux with a low-order adaptive
optics system, currently developed for the Calar Alto 2.2-m telescope. This could enable 
observations at shorter wavelengths or increase the resulting Strehl ratio under seeing conditions
normally unfavourable for Lucky Imaging. It has recently been proven at the 5-m Palomar telescope\cite{Law:2008}
that this approach actually works and can boost the performance of the Lucky Imaging technique.



\acknowledgments     
 
We are particulary grateful to Armin B\"ohm, Jens Helmling, and Uli Thiele
as well as all technical staff at Calar Alto and MPIA for their help in preparing and commissioning AstraLux. 

\bibliography{astralux}   
\bibliographystyle{spiebib}   

\end{document}